\documentclass[twocolumn]{aastex61}
\graphicspath{ {Figures/} }
\usepackage{amsmath}
\received{\today}
\revised{\today}
\accepted{\today}
\submitjournal{ApJ}

\shorttitle{}
\shortauthors{J.M.Cann et al}

\begin{document}

\title{Multi-wavelength observations of SDSS J105621.45+313822.1, a broad-line, low-metallicity AGN}

\correspondingauthor{Jenna M. Cann}
\email{jcann@masonlive.gmu.edu}

\author{Jenna M. Cann}
\affiliation{George Mason University, Department of Physics and Astronomy, MS3F3, 4400 University Drive, Fairfax, VA 22030, USA}
\affiliation{National Science Foundation, Graduate Research Fellow}

\author{Shobita Satyapal}
\affiliation{George Mason University, Department of Physics and Astronomy, MS3F3, 4400 University Drive, Fairfax, VA 22030, USA}

\author{Thomas Bohn}
\affiliation{Department of Physics and Astronomy, University of California, Riverside, 900 University Avenue, Riverside, CA 92521, USA}

\author{Remington O. Sexton}
\affiliation{Department of Physics and Astronomy, University of California, Riverside, 900 University Avenue, Riverside, CA 92521, USA}

\author{Ryan W. Pfeifle}
\affiliation{George Mason University, Department of Physics and Astronomy, MS3F3, 4400 University Drive, Fairfax, VA 22030, USA}

\author{Christina Manzano-King}
\affiliation{Department of Physics and Astronomy, University of California, Riverside, 900 University Avenue, Riverside, CA 92521, USA}

\author{Gabriela Canalizo}
\affiliation{Department of Physics and Astronomy, University of California, Riverside, 900 University Avenue, Riverside, CA 92521, USA}

\author{Barry Rothberg}
\affiliation{George Mason University, Department of Physics and Astronomy, MS3F3, 4400 University Drive, Fairfax, VA 22030, USA}
\affiliation{LBT Observatory, University of Arizona, 933 N. Cherry Ave., Tuscan, AZ 85721, USA}

\author{Mario Gliozzi}
\affiliation{George Mason University, Department of Physics and Astronomy, MS3F3, 4400 University Drive, Fairfax, VA 22030, USA}

\author{Nathan J. Secrest}
\affiliation{U.S. Naval Observatory, 3450 Massachusetts Avenue NW, Washington, DC 20392, USA}

\author{Laura Blecha}
\affiliation{University of Florida, Department of Physics, P.O. Box 118440, Gainesville, FL 32611-8440}



\begin{abstract}

In contrast to massive galaxies with Solar or super-Solar gas phase metallicities, very few Active Galactic Nuclei (AGN) are found in low-metallicity dwarf galaxies. Such a population could provide insight into the origins of supermassive black holes. Here we report near-infrared spectroscopic and X-ray observations of SDSS J105621.45+313822.1, a low-mass, low-metallicity galaxy with optical narrow line ratios consistent with star forming galaxies but a broad H$\alpha$ line and mid-infrared colors consistent with an AGN. We detect the [\ion{Si}{6}] 1.96$\mu$m coronal line and a broad Pa$\alpha$ line with a FWHM of $850 \pm 25$~km~s$^{-1}$. Together with the optical broad lines and coronal lines seen in the SDSS spectrum, we confirm the presence of a highly accreting black hole with mass $(2.2 \pm 1.3) \times 10^{6}$~M$_{\odot}$, with a bolometric luminosity of $\approx10^{44}$~erg~s$^{-1}$ based on the coronal line luminosity, implying a highly accreting AGN. Chandra observations reveal a weak nuclear point source with $L_{\textrm{X,2-10 keV}} = (2.3 \pm 1.2) \times 10^{41}$~erg~s$^{-1}$, $\sim 2$ orders of magnitude lower than that predicted by the mid-infrared luminosity, suggesting that the AGN is highly obscured despite showing broad lines in the optical spectrum. The low X-ray luminosity and optical narrow line ratios of J1056+3138 highlight the limitations of commonly employed diagnostics in the hunt for AGNs in the low metallicity low mass regime.

\end{abstract}




\section{Introduction} \label{sec:intro}

The vast majority of active galactic nuclei (AGNs) are found in massive bulge dominated galaxies with gas phase metallicities that are typically super solar \citep[e.g.,][]{storchibergmann1998,hamann2002}.  Since the gas phase metallicity is strongly correlated with the galaxy’s stellar mass, low metallicity AGNs are likely to reside in low mass galaxies. The hunt for AGNs in dwarf galaxies has been an active field of research in recent years, since the black hole occupation fraction and mass distribution in the low mass regime place important constraints on models of supermassive black hole (SMBH) seed formation \citep[e.g.,][]{volonteri2009,volonteri2010, vanwassenhove2010, greene2012}. However, searches for AGNs in low mass galaxies have yielded only a small fraction of AGNs, all with solar or only slightly subs-solar metallicities.  This is a severe limitation,
since the premise behind the use of dwarf galaxies to probe seed black holes rests on the assumption that
they have had a quiescent cosmic history, free of external factors such as merging or tidal stirring, both of
which would drive gas to the center, fueling star formation, enriching the gas, growing a bulge,
and potentially fueling the SMBH. For example, in an extensive search of type 2 AGNs using the Sloan Digital Sky Survey (SDSS)  \citet{groves2006} and \citet{barth2008} found only a small fraction of AGNs residing in low mass hosts, all with either Solar or only slightly sub-Solar metallicities.  Similarly, in the recent optical survey of dwarf galaxies by \citet{reines2013}, nearly all of the dwarf galaxies with narrow emission line ratios consistent with AGNs have at least Solar metallicities, consistent with the high metallicity range found for the low mass broad line AGNs identified by \citet{greene2007}. While extremly rare, there have been a few low metallicty AGNs reported in the literature. \citet{izotov2007}, \citet{izotov2008}, and \cite{izotov2010} found evidence for extremely luminous broad line emission consistent with AGNs in a handful of low metallicity dwarf galaxies, most with optical narrow line ratios typical of HII regions, and \citet{schramm2013} found X-ray evidence for an AGN in several low mass galaxies, one of which is metal deficient, suggesting that low metallicity AGNs do exist. Of these few low metallicity AGNs, extensive multiwavelength observations are thus far lacking in the literature. 

\begin{figure}
\includegraphics[width=0.495\textwidth]{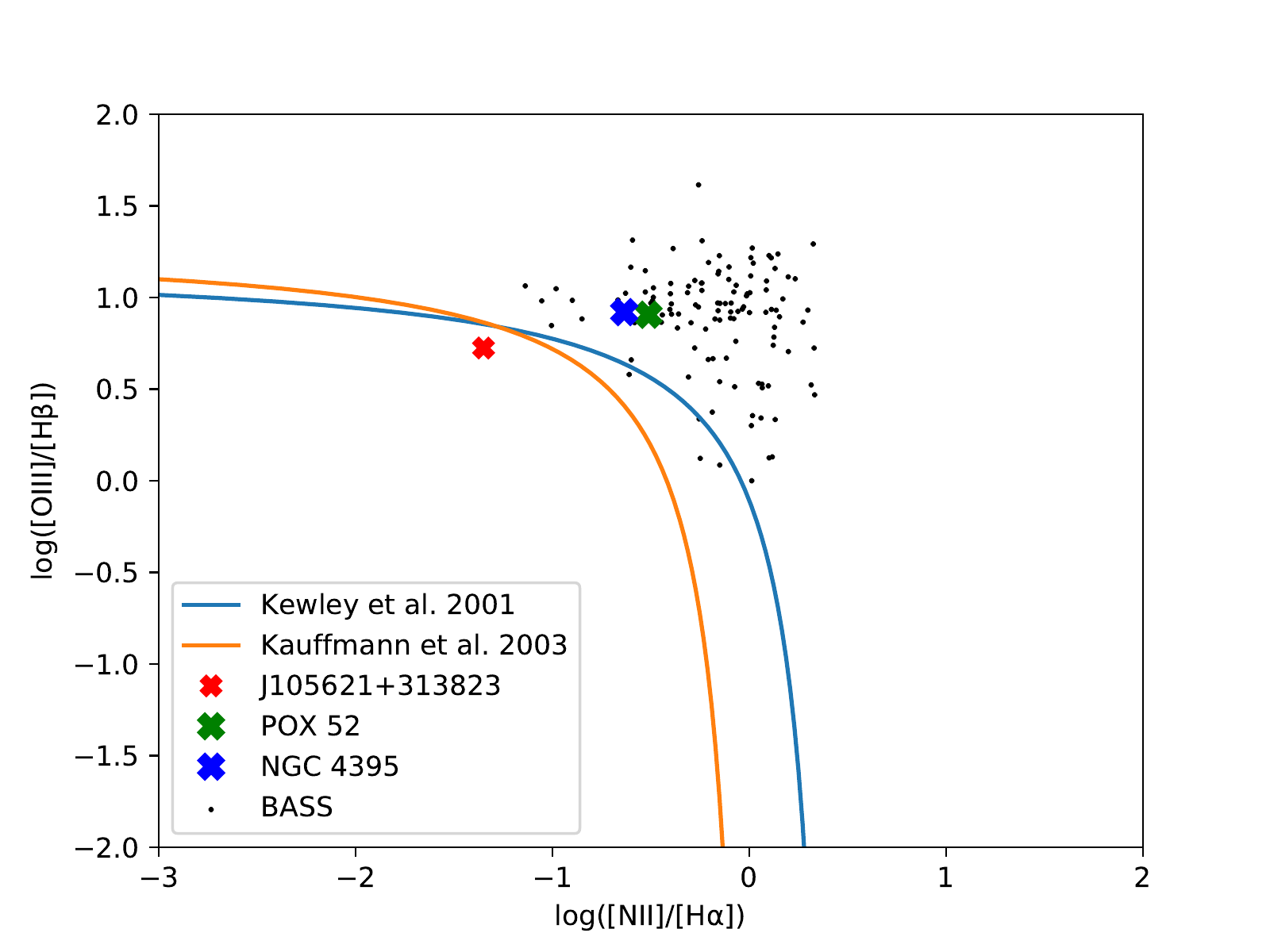}
\caption{BPT diagram comparing J1056+31 (red `x') with the BASS sample \citep{koss2017}, with the AGN \citep{kewley2001} and composite \citep{kauffmann2003} demarcation regions in blue and orange respectively.  Also plotted are NGC 4395 \citep[blue `x';][]{filippenko2003} and Pox 52 \citep[green `x';][]{barth2004}.  Note that J1056+31 is the only target that meets our metallicity selection criteria and displays star-forming ratios.}
\label{BPT}
\end{figure}

Here we present a multi-wavelength study of a low metallicity, broad line AGN, SDSS J105621.45+313822.1 (hereafter J1056+3138), a galaxy which is identified as a QSO broad line object by SDSS DR12, and in past catalogs has been classified as a broad line QSO \citep{dabrusco2009,richards2009,souchay2012,souchay2015,richards2015} or broad line AGN  \citep{toba2014,rakshit2017}. Based on the MPA/JHU catalog, the galaxy has a redshift of z=0.161 and a stellar mass of $10^{9.89}~\mathrm{M_{\odot}}$, roughly $2.5\times$ the mass of the Large Magellanic Cloud.  In Figure \ref{BPT}, we plot the location of J1056+3138 on the Baldwin-Phillips-Terlevich (BPT) diagram \citep{baldwin1981}. As can be seen, it has one of the lowest [\ion{N}{2}]/H$\alpha$ emission line ratios, a robust indicator of gas phase metallicity regardless of ionizing radiation field and ionization parameter \citep{groves2006}, compared to the entire sample of {\it Swift}/BAT AGNs from the 70 month catalog \citep{Baumgartner2013}, which comprise the most complete sample of hard X-ray (14 to 195 keV) selected AGNs in the local universe. Note that the well known dwarf galaxies with AGNs, NGC 4395 \citep{filippenko2003} and POX 52 \citep{barth2004}, both have at least $2-10\times$ higher gas metallicities than our target as suggested by their [N~II]/H$\alpha$ ratios shown in Figure \ref{BPT}. 

\begin{figure}
\includegraphics[width=0.495\textwidth]{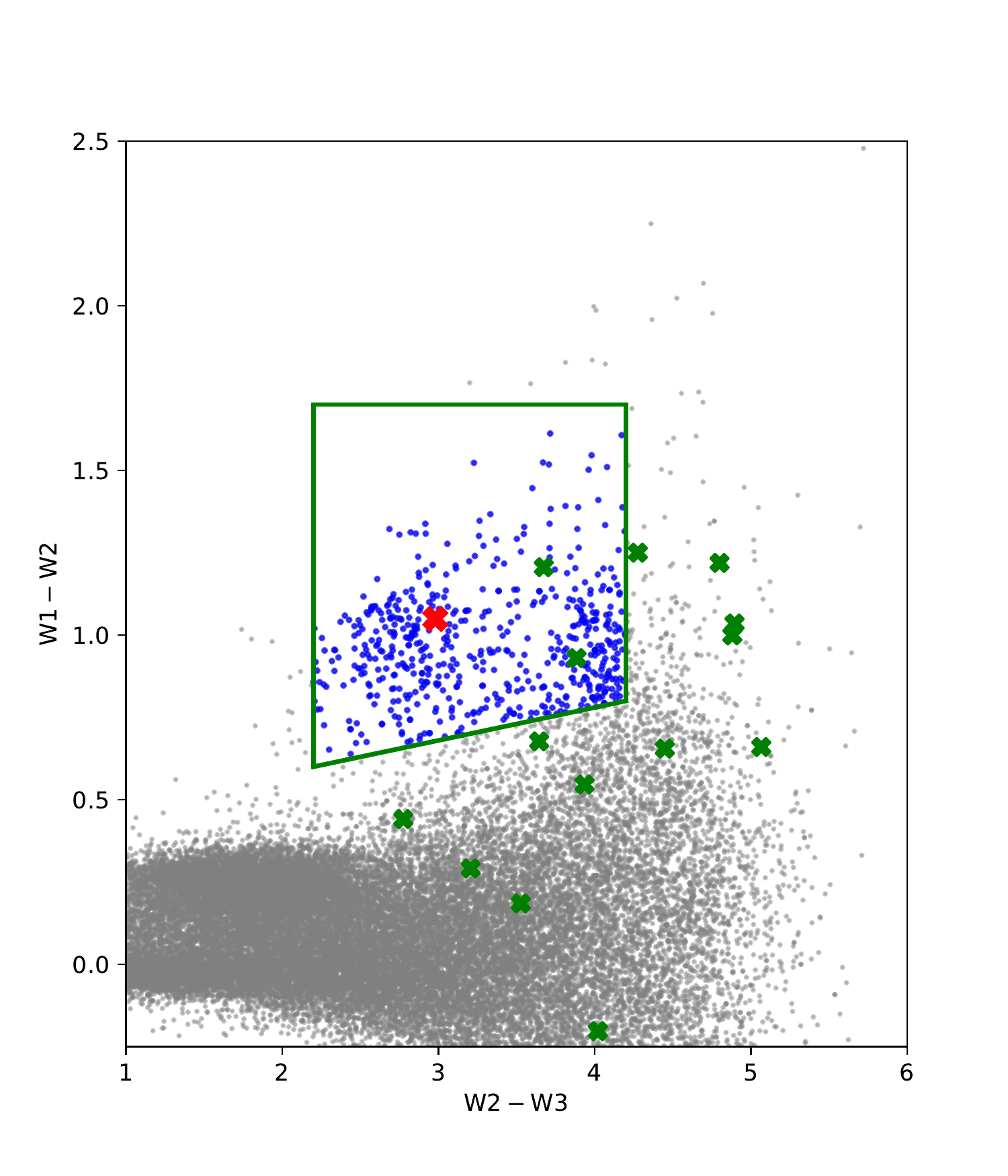}
\caption{Mid-infrared color-color diagram showing the placement of low metallicity galaxies with comparable $\log$[\ion{N}{2}]/H$\alpha$ to J1056+3138, as defined by [\ion{N}{2}]/H$\alpha \leq -1.3$ using emission line fluxes from the MPA catalog. The box corresponds to the \citet{jarrett2011} demarcation region.}  Grey points correspond to galaxies that would not be identified as potential AGN using the strict \citet{jarrett2011} color cut, and blue points denote galaxies that would be characterized as potential AGN.  The location of J1056+3138 is marked with a red `X' and locations of other published low metallicity galaxies are marked with green `X's \citep{thuan2005,izotov2007,izotov2008,izotov2012}.
\label{jarrett}
\end{figure}

In addition to the identification of broad lines, J1056+3138 displays mid-infrared colors suggestive of an AGN using the all-sky \textit{Wide-field Infrared Sky Explorer (WISE)} and the 3-band demarcation from \citet{jarrett2011}.  In general, low metallicity galaxies tend to be bluer, with relatively few displaying mid-infrared colors [3.4$\mu$m]-[4.6$\mu$m] (hereafter $W1-W2$) $ > 1.0$ \citep{griffith2011, izotov2011}.  Those that do show red \textit{WISE} colors tend to reside to the left of region that typically signifies dominant AGNs when plotted in $W1-W2$ vs. [4.6$\mu$m]-[12$\mu$m] (hereafter $W2-W3$) space \citep{jarrett2011}.  As can be seen in Figure \ref{jarrett}, only $\approx 0.7\%$ of galaxies with similar metallicity to J1056+3138 display mid-infrared colors characteristic of an AGN, further emphasizing the unique nature of this object.

Broad line AGNs with optical narrow line diagnostics consistent with star forming galaxies such as J1056+3138 constitute an extremely rare population. There are only $\approx 5\%$ of broad line AGNs falling in the star-forming region of the BPT diagram according to a recent study by \citet{stern2013}.  Using the full Max Planck Institut f\"{u}r Astrophysik/Johns Hopkins University (MPA/JHU) catalog\footnote{\url{http://www.mpa-garching.mpg.de/SDSS/}} of derived galaxy properties for the SDSS data release 8 (DR8), approximately $4\%$ of broad line galaxies have comparable comparable or lower [\ion{N}{2}]/H$\alpha$ emission line ratios indicating low gas phase metallicities. Only 1.5\% of all broad line AGNs have log([\ion{N}{2}]/H$\alpha$) ratios less than -1.3, the value for J1056+3138.


This paper is organized as follows.  In Section 2, we describe our near-infrared and X-ray observations and data analysis, followed by a description of our results in Section 3.  In Section 4, we discuss our results, and summarize our findings in Section 5.

We adopt a standard $\Lambda$CDM cosmology with $H_0=70$~km~s$^{-1}$~Mpc$^{-1}$, $\Omega_{\textrm{M}}=0.3$, and $\Omega_\Lambda=0.7$.

\section{Methodology} \label{sec:method}

\subsection{NIR Observations and Data Reduction}
We obtained near-infrared observations of J1056+3138 using the Near-Infrared Spectrograph \citep[NIRSPEC;][]{mclean1998} on the Keck II telescope on 2018 March 5, with a total exposure time of 32 minutes.  These observations were carried out using the low resolution mode with a slit width of $0.76\arcsec$ and a slit length of $42\arcsec$ to provide a resolution of $R \approx 1400$.  We used the filter NIRSPEC-7 for a wavelength coverage of $1.839-2.630~\mu$m.  Observations were done using an ABBA pattern, nodding along the slit.  This target was observed under clear weather conditions, with $\sim0.\arcsec5$ seeing.  A telluric standard (A0V) was observed immediately after at similar air mass. 

The data reduction was carried out using a modified version of LONGSLIT\_REDUCE\footnote{\url{http://www.astro.caltech.edu/~gdb/nirspec_reduce/nirspec_reduce.tar.gz}} for Keck NIRSPEC, and REDSPEC\footnote{\url{https://www2.keck.hawaii.edu/inst/nirspec/redspec.html}}.  These packages followed the standard steps for IR spectral data reduction, including flat fielding, sky subtraction, wavelength calibration, spectral extraction, and telluric correction.  Flux calibration was done in Python by estimating the flux of the telluric A0V star through its K-band magnitude and scaling the spectrum by this flux. A small corrective factor ($<5\%$) was included due to the wavelength difference between the center of K-band and that of the wavelength coverage used.  Line fluxes and uncertainties were determined from best-fit Gaussian models to the emission lines using a custom Bayesian maximum-likelihood code implemented in Python using the affine-invariant Markov Chain Monte Carlo (MCMC) ensemble sampler \textit{emcee} \citep{emcee}.

\subsection{Optical Observations and Analysis}
This object was observed twice by SDSS, once on 2004 May 12 with the SDSS spectrograph and once on 2013 March 18 with the BOSS spectrograph \citep{dawson2013}.  Aside from the emergence of a [\ion{Fe}{7}]5722 emission line in the 2013 that was not visible in 2004, these two spectra are nearly equivalent, with consistent broad and narrow line fluxes within photometric uncertainties, so the remainder of the paper will analyze the 2013 spectrum as it has the higher signal-to-noise.  We analyzed this spectrum in order to compare broad line fluxes and widths to our near-IR observations.  Optical spectral decomposition was performed on the SDSS spectrum using \textit{emcee} \citep{emcee} as done by \citet{sexton2019}.  For the H$\beta$/[\ion{O}{3}] region, we fit the region from restframe 4400 - 5800\AA, which includes the \ion{Mg}{1b} region used to estimate stellar velocity dispersion. All emission lines are modeled using Gaussians, with narrow-line FWHM and velocity offsets tied during the fitting process.  For the H$\alpha$/[\ion{N}{2}] region, we fit the region from restframe 6200 - 7000\AA.  The lack of prominent stellar absorption features in this region prevents us from using stellar template fitting, and thus the continuum is fit using only the power-law component, with amplitude and power-law index as free parameters.  
As with the near-IR data, line fluxes and uncertainties were determined using best-fit Gaussian models.

\subsection{X-ray Observations and Data Reduction}
\textit{Chandra} observations of this target were taken on 20 October 2019 in Cycle 19.  Data was taken with the ACIS-S instrument, with an exposure time of 16 ks, and pointed with the target centered at the aimpoint of the S3 chip.   The data was reduced and analyzed using the \textit{Chandra} Interactive Analysis of Observations (\textsc{ciao}) data package v4.11 and the Chandra Calibration Database v4.8.2 (\textsc{caldb}). After reprocessing and filtering the event file into full ($0.3-8$~keV), soft ($0.3-2$~keV), and hard ($2-8$~keV) energy bands, we used the \textsc{dmextract} module to extract the source counts from a 1.5\arcsec  radius aperture centered on the source position. Background counts were extracted from a 25\arcsec  radius aperture in the vicinity of the source and in an area free of other sources. Due to the low count nature of this source, we use binomial statistics, using the \citet{gehrels1986} approximation to account for the error in the source counts, where the upper bound is given by $1+\sqrt{x+0.75}$, the lower bound by $\sqrt{x-0.25}$, and x is the number of counts detected. All X-ray fluxes and luminosities quoted hereafter are derived from background-subtracted counts.

\section{Results}
\label{sec:results}
\subsection{Near-infrared and Optical Spectra of J1056+31}

\begin{figure*}[h]
\includegraphics[width=\textwidth]{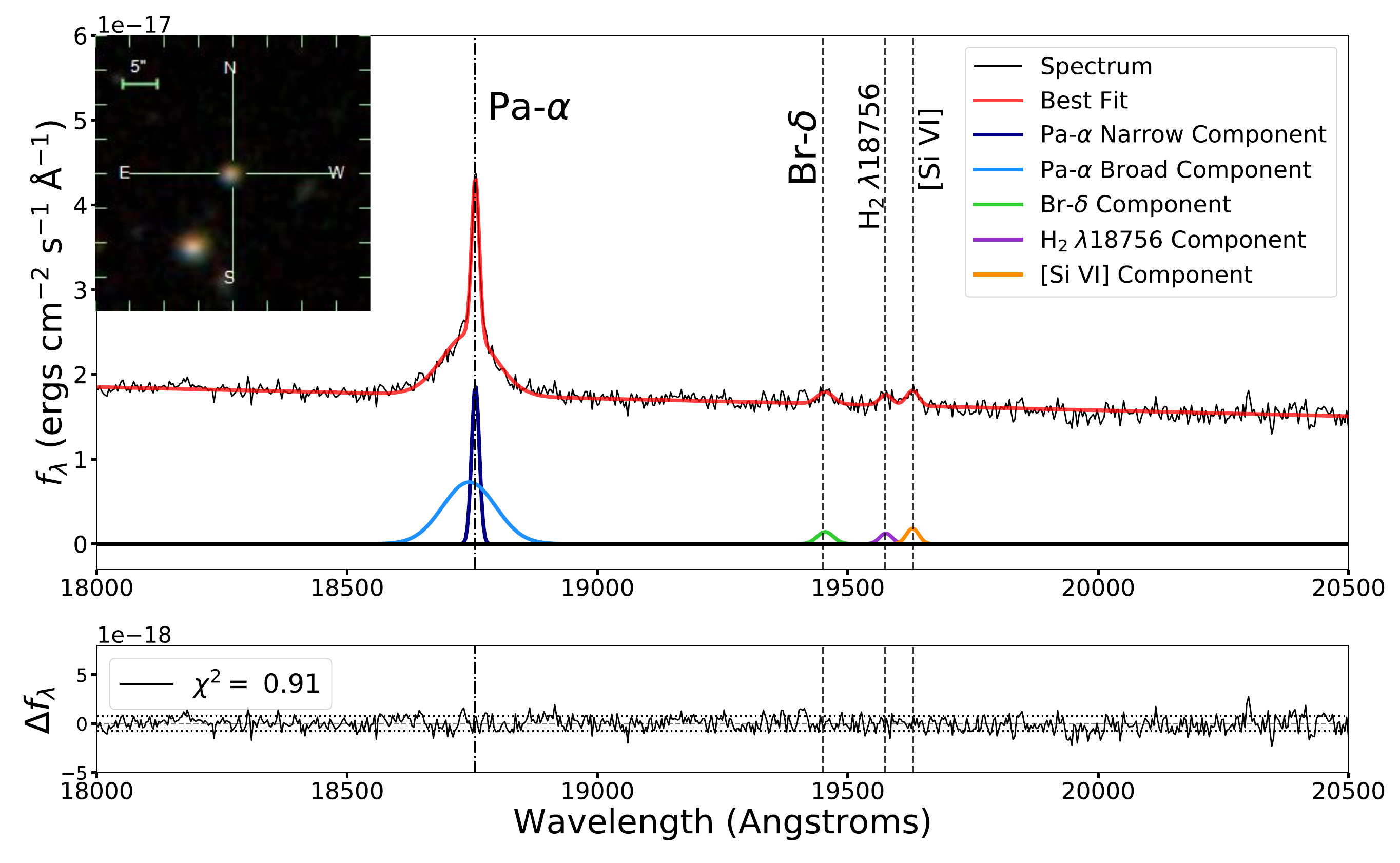}
\caption{K-band spectra from Keck NIRSPEC of J1056+3138. Note that this galaxy displays a prominent broad Pa$\alpha$ line and the [Si~VI] coronal line, as well as Br$\delta$ and H$_2$ lines. An SDSS \textit{gri} color composite image is also displayed, showing its compact morphology.}
\label{fig2}
\end{figure*}

\begin{figure*}[h]
\includegraphics[width=\textwidth]{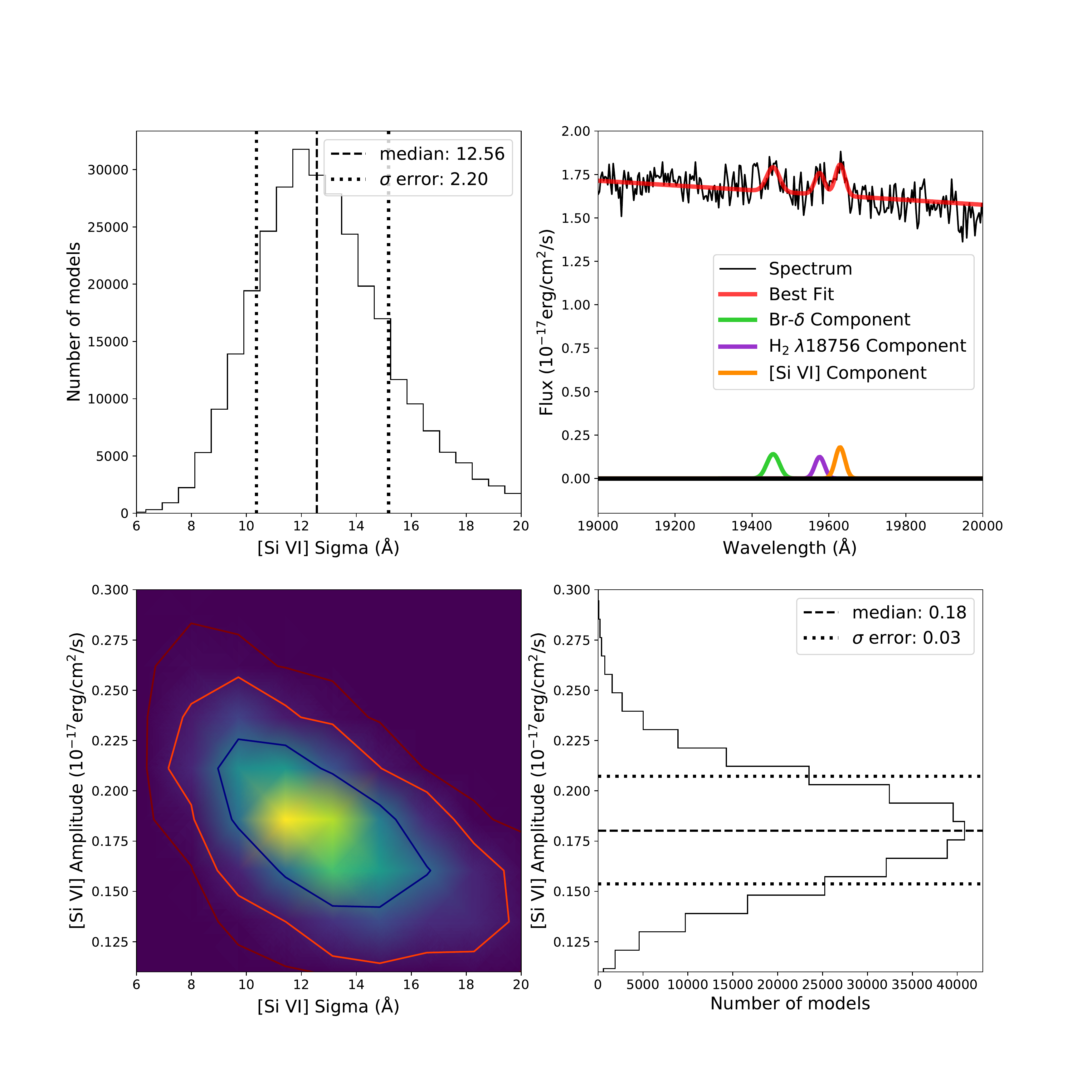}
\caption{Upper Right: The detected [\ion{Si}{6}] emission line, as well as nearby Br$\delta$ and H$_2$ lines. Lower Left: A heat map showing the output of the \textit{emcee} fits for the amplitude and width of the Gaussian for the [Si~VI] detection in J1056+3138. The contour levels show the $\sigma$ intervals up to 3$\sigma$, and the histograms (upper left and lower right) show the distribution of potential fits for the width (upper left) and amplitude (lower right) of the Gaussian.  As can be seen, the solution converged very well, proving a robust detection.}
\label{fig3}
\end{figure*}

The near-infrared  {\it K}-band spectrum of J1056+31 can be found in Figure \ref{fig2}. Most notably, this target contains a $3.3\sigma$ [\ion{Si}{6}] detection, with a flux of $(5.69\pm 1.75)\times10^{-17}~\mathrm{erg~cm^{-2}~s^{-1}}$ (Figure \ref{fig3}).  As \ion{Si}{6} has an ionization potential of 167 eV, even the hottest, most massive stars do not produce enough high energy ionizing radiation required to produce this ion (Satyapal et al. 2020, in prep). 

\begin{figure}
\centering
\includegraphics[width=0.4\textwidth]{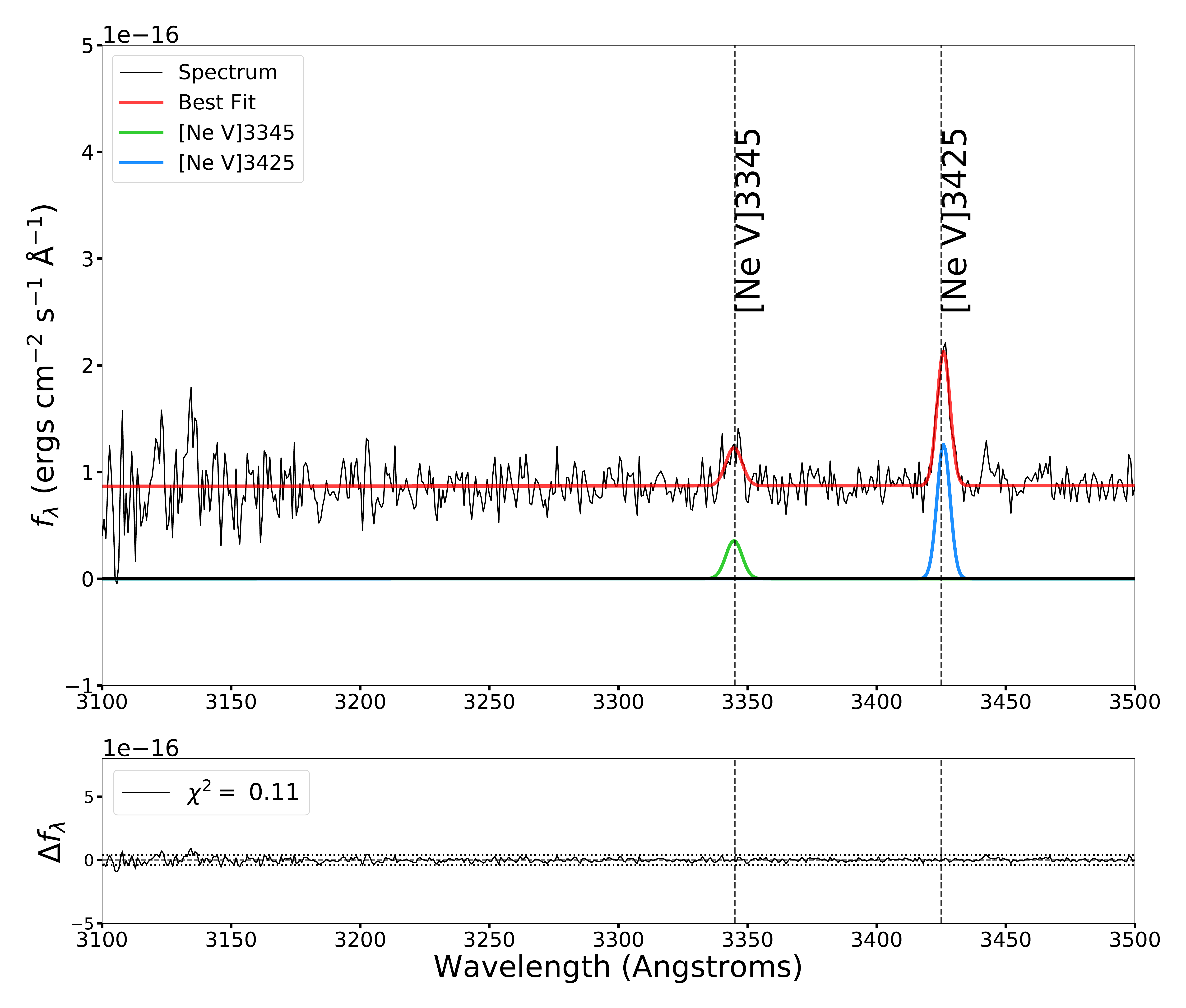}  \includegraphics[width=0.44\textwidth]{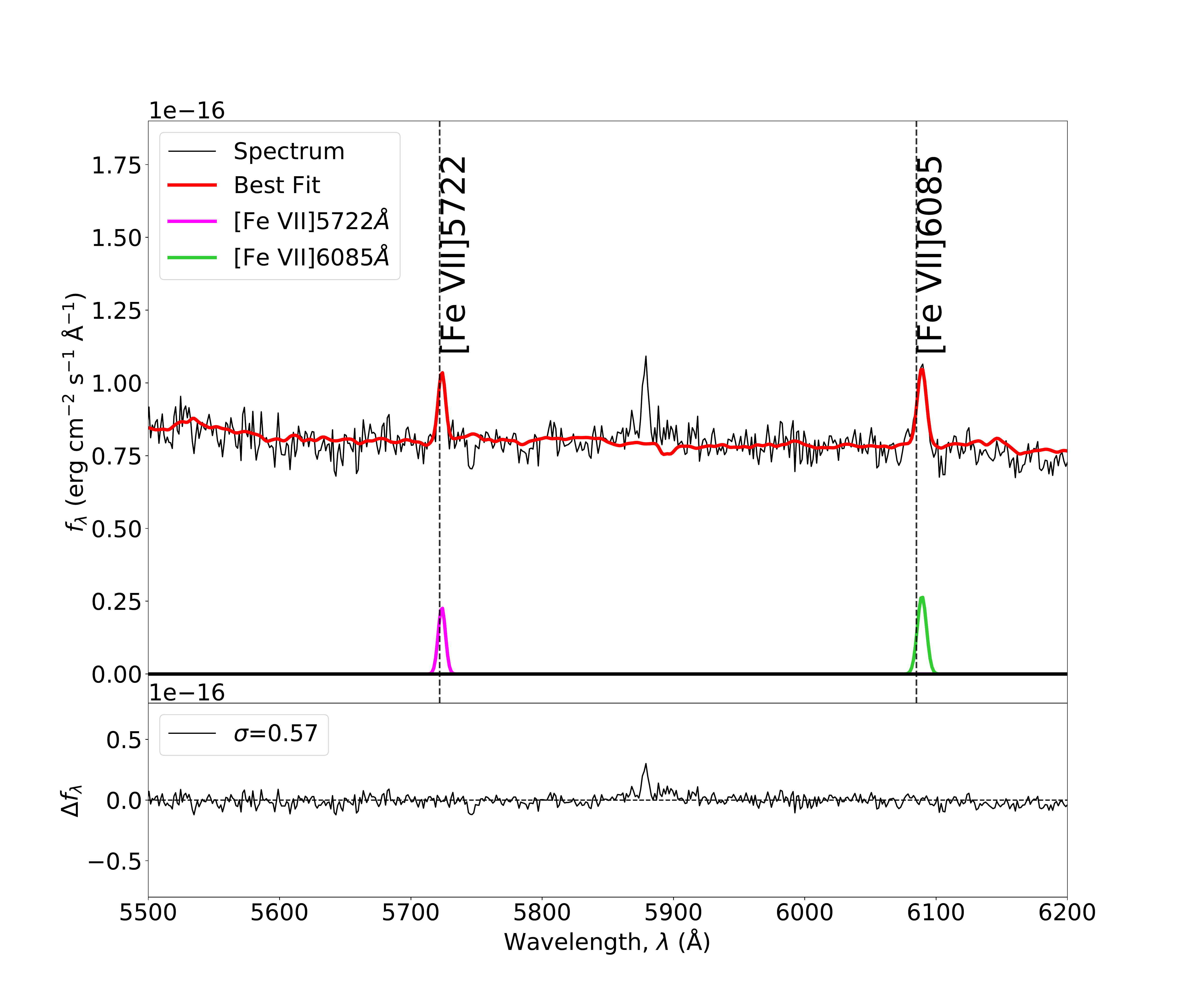}\\
\includegraphics[width=0.44\textwidth]{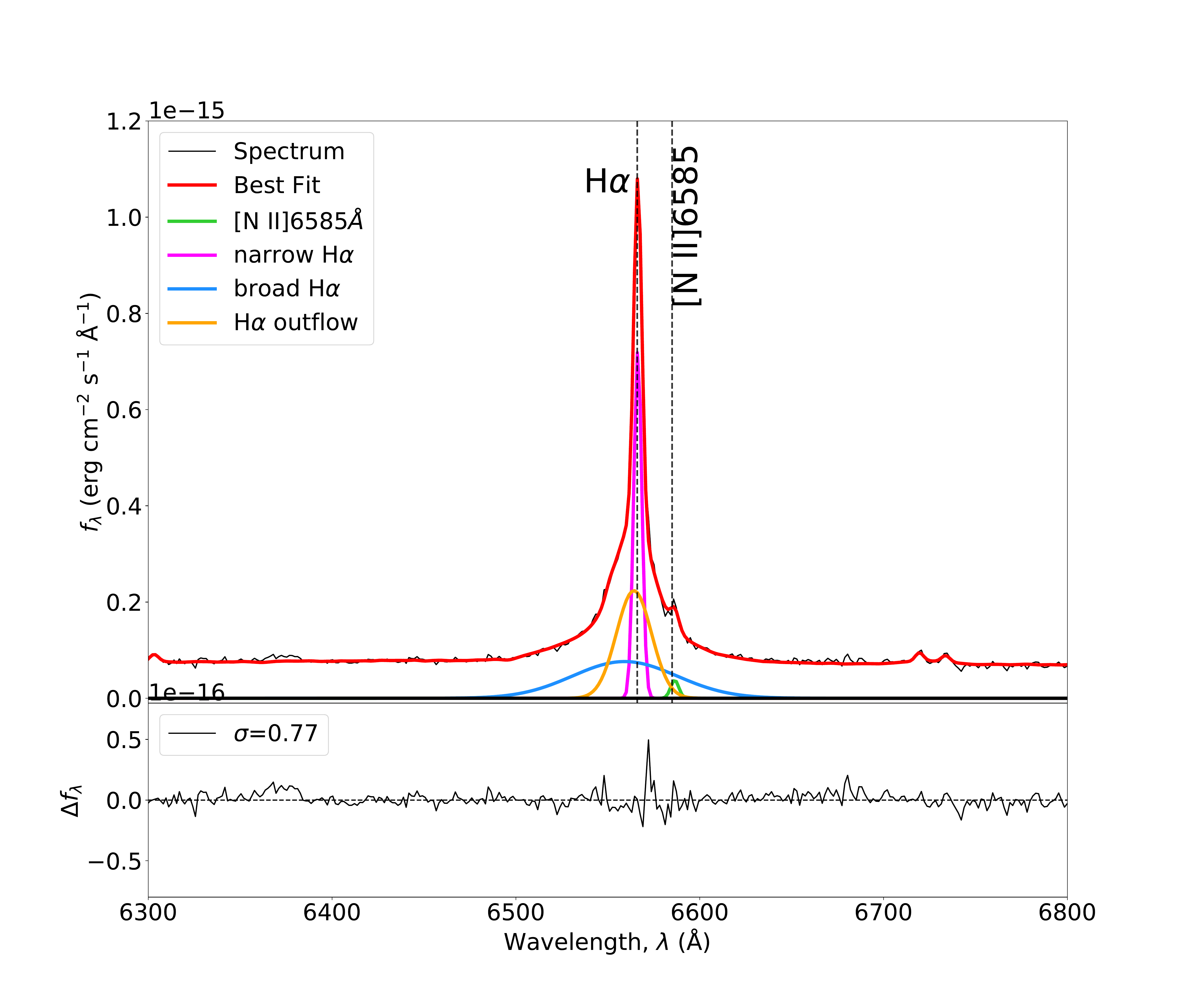}\\
\caption{Optical spectra from SDSS of J1056+3138.  The top panel displays the fits for the [\ion{Ne}{5}]$\lambda3345,3425$ coronal lines.  The center panel displays the fit for the [\ion{Fe}{7}]$\lambda5722,6085$ coronal lines.  The bottom panel displays the fit for the broad H$\alpha$.}
\label{fig4}
\end{figure}

The SDSS optical spectrum is shown in in Figure \ref{fig4}. Coronal lines were observed in both the SDSS and Keck observations.  The SDSS observations showed [\ion{Fe}{7}]$\lambda5722$ and [\ion{Fe}{7}]$\lambda6085$ emission lines, and  [\ion{Ne}{5}]$\lambda3425$ and [\ion{Ne}{5}]$\lambda3345$ lines.  These are the most ubiquitous optical coronal lines, which have been widely seen in AGN \citep[e.g.][]{appenzeller1991,vergani2018, yan2019}. Note that the ionization potentials of \ion{Fe}{7} and \ion{Ne}{5} are $99$~eV and $97$~eV, respectively, significantly less than the ionization potential of \ion{Si}{6}.  Fluxes for these lines can be found in Table \ref{tab1} for the [\ion{Si}{6}] and Table \ref{tab2} for the optical lines.  Note that the luminosity of the [\ion{Si}{6}] line is $3.18 \times 10^{39}$ erg s$^{-1}$, well within the range of luminosities of the well-studied AGN from \citet{mullersanchez2018,lamperti2017}, which range from $5.9 \times 10^{36} - 3.9 \times 10^{41}$ erg s$^{-1}$. The [\ion{Fe}{7}] and [\ion{Ne}{5}] lines in J1056+3138 have luminosities of $\approx 10^{40}$ erg s$^{-1}$, comparable to those observed in other AGN samples, with luminosities ranging from $10^{39} - 10^{42}$ erg s$^{-1}$ \citep{malkan1986,morris1988,storchibergmann1995}.

\begin{table*}[t]
\caption{Near-IR (Keck NIRSPEC) emission line fluxes}
\centering
\begin{tabular}{lcc}
\hline
\hline
\noalign{\smallskip}
        Line  & Wavelength & Flux \\
        & \AA & $10^{-17}~\mathrm{erg~cm^{-2}~s^{-1}}$\\
        \hline
        Pa$\alpha$ & 18756 & $35.90 \pm 1.73$ \\
        \hline
        br. Pa$\alpha$ & 18756 & $96.72 \pm 5.78$ \\
        \hline
        Br$\delta$ & 19451 & $5.60 \pm 1.85$ \\
        \hline
        H$_2$ & 19576 & $3.86 \pm 1.28$ \\
        \hline
        [\ion{Si}{6}] & 19628 & $5.69 \pm 1.75$ \\

    \noalign{\smallskip}
    \hline
    \noalign{\smallskip}       
     \end{tabular}
     \tablecomments{Broad lines are identified by ``br."}
     \label{tab1}
 \end{table*}

As can be seen in Figure \ref{fig2}, the spectra also contains a broad Pa$\alpha$ line with a flux of $(35.90 \pm 1.73) \times 10^{-17}$ erg cm$^{-2}$ s$^{-1}$ and a width of $850 \pm 25$ km s$^{-1}$. A Br$\delta$ and H$_2$ line were also detected. Fluxes for all infrared lines can be found in Table \ref{tab1}.  While its optical narrow emission lines placed the object in the ``star-forming" region of the BPT diagram (see Figure \ref{BPT}), J1056+3138 showed a broad H$\alpha$ line, as well as [Fe~VII]$\lambda5722,6085$ and [Ne~V]$\lambda3425,3345$ coronal lines, as can be seen in Figure \ref{fig4}.  While broad lines and coronal lines can be indicative of an AGN, their presence in a BPT star-forming galaxy can often be due to supernova activity, which fades over time \citep{baldassare2016}.  The first observation of J1056+3138 was taken on 2004 May 12, and our Keck observations took place on 2018 Mar 5.  There is a 14 year baseline between these two observations, so we can explore if there is any fading of the broad lines, which would be indicative of a stellar origin to the broad line rather than an AGN.  We compared the extinction corrected optical broad line flux to the broad Pa$\alpha$ flux.  The theoretical H$\alpha$/Pa$\alpha$ ratio, assuming Case B recombination, is 8.5 \citep{osterbrock2006}.  The observed value is $6.0 \pm 0.1$, implying negligible extinction toward the ionized gas, and demonstrating that there is no fading of the recombination line flux over a 14 year baseline, ruling out the possibility that the broad lines are due to supernova activity.  Since we have multiple recombination lines, we also estimated the extinction using the Pa$\alpha$ and Br$\delta$ lines.  Using the observed ratio of these fluxes, we find an $A_\mathrm{V} < 1$, assuming a Milky Way-like extinction curve ($R\mathrm{_{V}}$=3.1), again implying that there is no fading in recombination line fluxes between the SDSS and near-infrared observations.

\begin{table*}[t]
\caption{Optical SDSS emission line fluxes}
\centering
\begin{tabular}{lcc}
\hline
\hline
\noalign{\smallskip}
        Line  & Wavelength & Flux \\
        & \AA & $10^{-17}~\mathrm{erg~cm^{-2}~s^{-1}}$\\
        \hline
        [\ion{Ne}{5}] & 3345 & $28.2 \pm 11.4$ \\
        \hline
        [\ion{Ne}{5}] & 3426 & $79.2 \pm 9.9$ \\
        \hline
        [\ion{O}{3}] & 4363 & $65.2 \pm 2.8$\\
        \hline
        H$\beta$ & 4861 & $82.7 \pm 3.4$ \\
        \hline
        br. H$\beta$ & 4861 & $281.6 \pm 9.0$ \\
        \hline
        [\ion{O}{3}] & 4959 & $135.0 \pm 4.9$ \\
        \hline
        [\ion{O}{3}] & 5007 & $407.4 \pm 14.7$ \\
        \hline
        [\ion{Fe}{7}] & 5722 & $16.0 \pm 1.8$ \\
        \hline
        [\ion{Fe}{7}] & 6085 & $23.9 \pm 2.3$ \\
        \hline
        [\ion{O}{1}] & 6300 & $10.3 \pm 3.0$\\
        \hline
        [\ion{N}{2}] & 6549 & $7.0 \pm 0.7$\\
        \hline
        H$\alpha$ & 6563 & $404.5 \pm 10.8$ \\
        \hline
        br. H$\alpha$ & 6563 & $578.3 \pm 13.1$ \\
        \hline
        [\ion{N}{2}] & 6583 & $20.5 \pm 2.1$ \\
        \hline
        [\ion{S}{2}] & 6717 & $10.8 \pm 1.8$ \\
        \hline
        [\ion{S}{2}] & 6730 & $7.6 \pm 1.4$ \\
        
    \noalign{\smallskip}
    \hline
    \noalign{\smallskip}       
     \end{tabular}
    \tablecomments{Broad lines are identified by ``br."}
     \label{tab2}
 \end{table*}
 




\subsection{X-ray Results}

We detected an X-ray point source coincident with the SDSS optical source with an apparent luminosity, uncorrected for intrinsic absorption, of $L_{\textrm{X, 2-10 keV}} = (2.3 \pm 0.9) \times 10^{41}$ erg~s$^{-1}$, with six counts combined in both hard and soft bands.  Due to the low number of counts, we explored the binomial no-source probability of the detection, $P_{\rm{B}}$, which is proportional to the probability that the measured counts are due to spurious background activity (see \citealp{weisskopf2007,lansbury2014} for the mathematical expression for this statistic). Adopting the requirement that real sources (not due to background activity) satisfy a threshold of $P_{\rm{B}}<0.002$ \citep{satyapal2017}, and noting that the X-ray source yields a no-source probability of $\log(P_{\rm{B}})\sim-6.4$, we conclude that the X-ray source is unlikely to be due to spurious background activity.  The typical $L_{\textrm{2-10keV}}$ threshold generally adopted by the community to unambiguously identify an AGN is $10^{42}$ erg~s$^{-1}$ \citep{zezas2001,ranalli2003,wang2013}. While the X-ray luminosity for J1056+3138 is somewhat higher than seen in star forming galaxies, \citet{brorby2014} find that the X-ray emission produced for a given SFR is approximately an order of magnitude larger than that found in near solar metallicity galaxies (see also \citet{kaaret2011}).  Further, \citet{prestwich2013} also show that ULXs are more common in low metallicity systems, adding further to the ambiguity of the origin of the X-ray emission in J1056+3138 and making an AGN identification by this observed X-ray luminosity alone tentative.

The observed luminosity is two orders of magnitude lower than what is expected from the $L_X - L_{12\mu \textrm{m}}$ relation ($\approx10^{43}$ erg~s$^{-1}$) \citep{secrest2015}.  Taking the relation between [\ion{Si}{6}] and $L_X$ from \citet{lamperti2017}, an estimate of $L_X \approx 10^{43}$ is calculated, also two orders of magnitude above the detected luminosity.  These discrepancies strongly suggest that the source is heavily obscured along the line of sight. An estimate for $N_H$ was calculated based on the relation determined in Pfeifle et al. 2020 (in prep) using 12$\mu$m luminosity data from \textit{WISE} and X-ray data from the \textit{BASS} Survey, which provides $N_{H} \approx (7 \pm 2) \times 10^{24}$~cm$^{-2}$.

\begin{figure}
\centering
\includegraphics[width=0.44\textwidth]{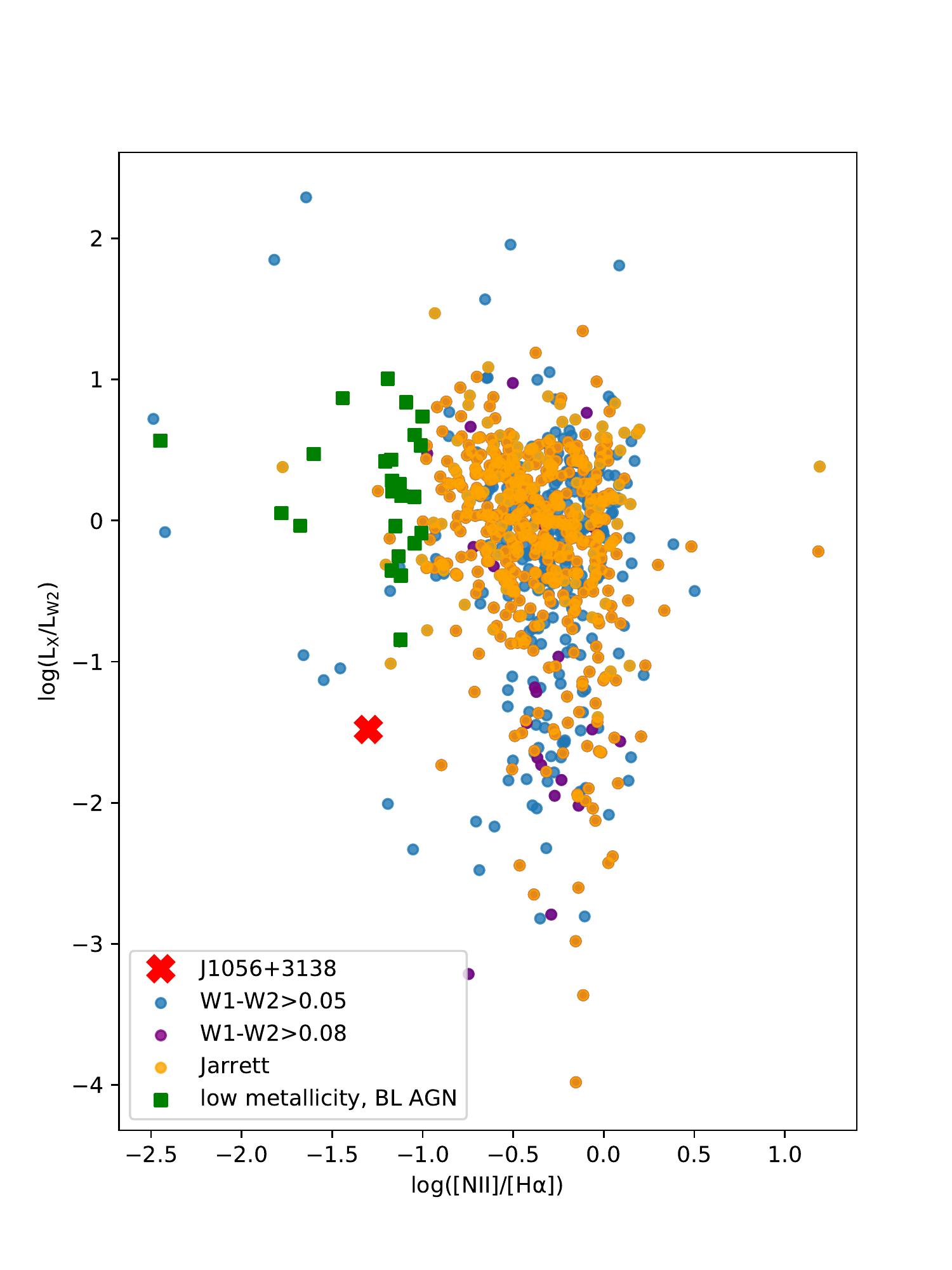}
\caption{$L_{X}/L_{W2}$ compared with metallicity diagnostic [\ion{N}{2}]/H$\alpha$] for galaxies in the MPA-JHU catalog (DR8) that show mid-infrared colors suggestive of AGN activity based on three color cuts: W1-W2 $>0.5$ (blue), W1-W2 $>0.8$ (purple), and the \citet{jarrett2011} cut (orange).  Galaxies with optical broad lines and [\ion{N}{2}]/H$\alpha$ ratios that fall below the low metallicity cut-off ($\log$([\ion{N}{2}]/H$\alpha$)$< -1.0$) are shown as green squares, and J1056+3138 is shown as a red `X'.  Note that J1056+3138 has the lowest $L_X/L_{W2}$ ratio when compared to similar galaxies.}
\label{fig5}
\end{figure}

We compared this discrepancy between the X-ray and mid-IR luminosities to other similar targets in the MPA-JHU catalog in Figure \ref{fig5}, where we've plotted the $L_X/L_{W2}$ of various mid-IR selected AGNs compared to broad line AGNs with low metallicity and J1056+3138 compared with their $\log$([\ion{N}{2}]/H$\alpha$).  Of the low metallicity galaxies with broad line AGNs, J1056+3138 has the lowest X-ray luminosity when compared to its W2 luminosity.  This apparent deficit in X-ray radiation is in agreement with other X-ray studies done on low metallicity galaxies, where it is found that X-ray luminosities tend to be one to two orders of magnitude below that expected from multi-wavelength diagnostics, due either to obscuration of the X-ray emission or an intrinsic X-ray weakness \citep{simmonds2016}.  This could possibly be due to the lack of an emitting corona, which may be characteristic of a few broad absorption line (BAL) quasars and ultra-luminous infrared galaxies \citep{luo2014, teng2015}.  In addition for low-mass cases, \citet{dong2012b} find several AGNs in low mass galaxies that appear to be X-ray weak, possibly due to potential changes in the accretion disk temperature or structure which would impact the fraction of the disk energy that is reprocessed into a corona.  This is supported by the low $L_{\textrm{X,2-10 keV}}$ vs. $L_{\textrm{[O~III]}}$ ratio of several low mass galaxies in their sample.  In J1056+3138, however, the [O~III] emission is dominated by star formation, complicating the interpretation of the relationship between these luminosities in this source.  It is also possible that there is a different dust-to-gas ratio in this galaxy, potentially affecting the relationship between X-ray absorption and extinction by dust of the broad line region 
\citep{groves2006}. Further, as the \textit{WISE} and \textit{Chandra} observations were not simultaneous, with a 9 year separation between the observations, it is also possible that variability could play a role in this discrepancy, although based on typical flux changes, it may not be able to account for the anomalous ratio found for J1056+3138 \citep{maughan2019,sheng2017}.  Due to its low X-ray luminosity, whether due to obscuration, variability, or an inherent X-ray weakness, J1056+3138 would not have been confirmed as an AGN through X-ray follow-up, showing the necessity for multi-wavelength studies when dealing with low mass and low metallicity galaxies.  If even a very powerful AGN can be hidden in the X-rays, many more less powerful and low luminosity AGNs may be hiding inside low metallicity dwarf galaxies.

\subsection{Black Hole Mass and Luminosity Estimates}

Black hole masses were calculated using the widths of the broad H$\alpha$ \citep{woo2015} and Pa$\alpha$ \citep{kim2018} lines. The broad Pa$\alpha$ has a width of $850 \pm 25$ km~s$^{-1}$, corresponding to a black hole mass of $(2.2 \pm 1.3) \times 10^6$ M$_{\odot}$, or $10^{6.4}$M$_{\odot}$, which was obtained with Equation 10 in \citet{kim2018}.  In this relation, they used the new virial factor $\log f = 0.05 \pm 0.12$ that was derived in \citet{woo2015}.  The broad H$\alpha$ has a width of $1129 \pm 21$ km s$^{-1}$, corresponding to a black hole mass of $(3.4 \pm 1.4) \times 10^6$ M$_{\odot}$, or $10^{6.5}$M$_\odot$, which was obtained used Equation 5 in \citet{woo2015}.

We estimated the bolometric luminosity of $\approx10^{44}~\mathrm{erg~s^{-1}}$, however there is considerable scatter in the $L\mathrm{_{[Si~VI]}}$ vs. $L\mathrm{_{14-195 keV}}$ relation reported in \citet{lamperti2017} with a number of upper limits in their sample.  We used a $L\mathrm{_{14-195 keV}}$/$L\mathrm{_{bol}}$ factor as reported in \citet{winter2012}.  We used the [\ion{Si}{6}] to estimate a bolometric luminosity since the [\ion{O}{3}] emission is likely dominated by star formation and cannot be used to obtain a bolometric luminosity as is often done in optically identified AGNs.  This luminosity, along with the Eddington luminosities calculated from the mass estimates, implies a high Eddington rate, $L/L_{\textrm{Edd}}\approx0.3$.  These Eddington ratios are similar to those reported in \citet{greene2007} and suggest that this is a highly accreting black hole.

\section{Discussion}
\label{sec:discussion}

\subsection{Coronal Lines}

While coronal lines have often been used as a robust AGN indicator, there is the possibility that these lines arise in star forming regions. In principle, Wolf-Rayet stars and shock excitation in starburst driven winds can generate broad lines and even high ionization lines \citep{schaerer1999,abel2008,allen2008}. Since the hardness of the stellar radiation field increases with decreasing metallicity \citep{campbell1986}, enhanced emission from lines corresponding to higher ionization potentials is expected for metal deficient galaxies. High ionization lines have indeed been detected in H~II regions in blue compact dwarfs and planetary nebulae \citep{feibelman1996,fricke2001,thuan2005,izotov2012} with a detection rate that appears to be correlated with decreasing metallicity \citep{izotov2012}. However, the line luminosities are weaker than those found in AGN, by up to four orders of magnitude \citep{izotov2012}, and the emission line fluxes are weak compared to the recombination line fluxes.  For example, \citet{thuan2005} and \citet{izotov2012} finds [\ion{Ne}{5}] emission in ten HII regions in blue compact dwarfs (BCDs), with ratios that range from $0.005-0.03$ for [\ion{Ne}{5}]$\lambda3345$/H$\beta$ and from $0.003-0.005$ for [\ion{Ne}{5}]$\lambda3425$/H$\beta$, 3 orders of magnitude below what is found in J1056+3138.  \citet{izotov2001,izotov2004} also find [\ion{Fe}{7}] emission in two blue compact dwarf galaxies, with ratios of $\approx 0.001$ for [\ion{Fe}{7}]$\lambda5722$/H$\beta$ and $\approx 0.0002-0.002$ for [\ion{Fe}{7}]$\lambda6085$/H$\beta$, 2-3 orders of magnitude below what is found in J1056+3138.  Note that the [\ion{Fe}{7}] and [\ion{Ne}{5}] lines in J1056+3138 have luminosities of $\approx 10^{40}$ erg s$^{-1}$, comparable to those observed in other AGNs, with luminosities ranging from $10^{39} - 10^{42}$ erg s$^{-1}$ \citep{malkan1986,morris1988,storchibergmann1995}, as is the case for the [\ion{Si}{6}] line as discussed in Section \ref{sec:results}. Moreover, while the optical coronal lines have been detected in H~II regions and BCDs that may not host AGNs, the ionization potential of \ion{Si}{6} (167 eV) is significantly higher than that of \ion{Fe}{7} or \ion{Ne}{5} and is only created in very extreme conditions when not in the presence of an AGN.  Note that this line has been seen in planetary nebulae and Galactic supernovae, but its luminosity is eight orders of magnitude lower than that of J1056+3138 and other AGNs \citep{ashley1988,benjamin1990,greenhouse1990}, making these sources undetectable outside of the Milky Way.  Thus far, there is no evidence of any coronal line with ionization potential above $\approx$ 100 eV in a purely star-forming galaxy, making it highly implausible that star formation is responsible for the coronal lines detected in J1056+3138.

\subsection{Metallicity Estimates}

Current studies of metallicity estimation assume an SED that is either primarily AGN or stellar in origin.  As J1056+3138 hosts an AGN, yet displays star-forming colors and has a strong stellar component, determining a precise SED, and thus a precise estimate of abundances is extremely difficult.  However, an initial estimate is calculated using the diagnostics determined by photoionization simulations and tested on low metallicity dwarf galaxies, including those with AGNs \citep{izotov2006,izotov2007,izotov2008}.  This metallicity estimate uses variables $t$, temperature $T_e(\mathrm{O~III})$, $C_T$, and $x$ as defined by

\begin{equation}
    t = \frac{1.432}{\log{[\lambda4959 + \lambda5007/\lambda4363]}-\log C_T}
\end{equation}

\begin{equation}
    t = 10^{-4}T_e(\mathrm{O~III})
\end{equation}

\begin{equation}
    C_T = (8.44-1.09t+0.5t^2-0.08t^3)\frac{1+0.0004x}{1+0.044x}
\end{equation}

\begin{equation}
    x = 10^{-4}N_et^{-0.5}
\end{equation}

where the density, $N_e$, is 10 \citep{izotov2007}, as well as emission lines [\ion{O}{3}]5007, [\ion{O}{3}]4959, [\ion{O}{2}]3726, and H$\beta$. The abundances are then derived using the following:

\begin{equation}
\begin{split}
    12 + \log\mathrm{O^+/H^+} = \log{\frac{\lambda3727}{\mathrm{H}\beta}} + 5.961 + \frac{1.676}{t}\\
    - 0.40 \log t - 0.034t + \log(1+1.35x)
\end{split}
\end{equation}

\begin{equation}
\begin{split}
    12 + \log\mathrm{O^{2+}/H^+} = \log\frac{\lambda4959 + \lambda5007}{\mathrm{H}\beta} + 6.200 \\
    + \frac{1.251}{t} - 0.55 \log{t} - 0.014 t
\end{split}
\end{equation}

Considering a value of 12 + $\log$(O/H) of 8.69 for Solar \citep{groves2006,asplund2006}, these relations provide a metallicity estimate for J1056+3138 of approximately 10\% Solar.  This is in agreement with our initial [\ion{N}{2}]/H$\alpha$ cutoff criteria.

\subsection{Implications}
The multi-wavelength study of J1056+3138 has broad astrophysical implications for our understanding of the origins of supermassive black holes. J1056+3138 is in one of the lowest metallicity galaxies known to contain an AGN, and it is one of the lowest metallicity, low mass galaxies to show a high ionization infrared coronal line.  While it is a broad line AGN with strong coronal lines in the optical, its BPT line ratios suggest the dominant emission is stellar in origin.  While mid-infrared color selection picks it out as a dominant AGN, with a bright Pa$\alpha$ and strong [\ion{Si}{6}] coronal line in its K band spectrum, it is barely visible when searching for an X-ray point source, and it would easily be mistaken for X-ray binaries, if it was detected at all.  The results presented here strongly support that low mass accreting black holes exist in galaxies that show no evidence for AGNs using traditional diagnostics, such as X-ray detections and BPT line ratios, calling into question the current occupation of fraction of AGNs in the low mass and low metallicity regime and highlighting the importance of multi-wavelength studies to obtain a complete census of AGNs in the low mass, low metallicity regime. 

Using SDSS and the \textit{WISE} survey, there are $\sim1500$ low mass ($M_* < 10^{10.5}$M$_{\odot}$), low metallicity ($\log$(\ion{N}{2}/H$\alpha$)) galaxies with infrared colors suggestive of AGN, many of which have stellar masses as low as $10^6 M_{\odot}$.  These are prime candidates for follow-up with the \textit{James Webb Space Telescope (JWST)} to search for accreting intermediate mass black holes. None of these galaxies show optical emission line ratios indicative of AGNs, underscoring the limitations of optical studies in the search for accreting black holes in galaxies that may be truly more representative of the local analogs of early galaxies.  Currently, there is no direct evidence for black holes with masses anywhere between $\approx100-10,000~\mathrm{M_{\odot}}$. This technique can be a powerful avenue in which black holes in this ``mass desert'' can finally be discovered using future high sensitivity \textit{JWST} observations \citep{cann2018}.  While coronal lines were also seen  in the SDSS spectrum of J1056+3138, the predicted fluxes of optical coronal lines are a factor of at least 5 times less than the infrared coronal lines even in the absence of any extinction (Cann et al. 2020, in prep).  The infrared coronal line fluxes are also enhanced when the black hole mass decreases, highlighting the need for infrared spectroscopic observations in the hunt for intermediate-mass black holes \citep{cann2018}.

\section{Conclusions} \label{sec:conclusions}
We present a multi-wavelength study of J1056+3138, a low metallicity, broad line AGN, including observations from \textit{Chandra}, \textit{Keck}/NIRSPEC, \textit{WISE} and \textit{SDSS}.  Our main results can be summarized as follows:

\begin{enumerate}
    \item In its \textit{SDSS} spectrum, J1056+3138 displays optical emission line ratios suggestive of a purely star-forming galaxy, despite strong optical broad lines and four optical coronal lines, [\ion{Ne}{5}]3345,3425 and [\ion{Fe}{7}]5722,6085. 
    \item Observations with \textit{WISE} show J1056+3138 to have mid-infrared colors suggestive of a strong AGN, despite only $\approx 0.7\%$ of similarly low metallicity galaxies residing in the same color-color space.
    \item K-band observations with \textit{Keck}/NIRSPEC revealed a broad Pa$\alpha$ line that, when its width and luminosity are compared to that of the broad H$\alpha$, implies negligible extinction.  These observations also showed a [\ion{Si}{6}] coronal line, making J1056+3138 the lowest metallicity galaxy to show this line.
    \item \textit{Chandra} observations detected an X-ray point source coincident with the SDSS optical source, but at a luminosity of only $\approx 10^{41}$ erg~s$^{-1}$, two orders of magnitude below that expected based on its mid-infrared luminosity.  This discrepancy imples the source is either heavily obscured, with an $N_H \approx (7 \pm 2) \times 10^{24}$ cm$^{-2}$, or inherently X-ray weak.
    \item Black hole mass estimates were calculated using the widths of the broad H$\alpha$ and Pa$\alpha$ lines.  Estimates ranged from $2.2 - 3.4 \times 10^6$ M$_\odot$, implying this AGN is fairly massive despite its low metallicity.
\end{enumerate}

Our results highlight the need for a multi-wavelength approach to truly characterize the source of ionization in this population.\\




\acknowledgments
J.M.C. gratefully acknowledges support from an NSF GRFP, a Mason 4-VA innovation grant, a Cosmos Club Foundation Cosmos Scholar grant, a Sigma Xi Grant-in-Aid of Research \#G2018031594610583. S.S. and J.M.C. acknowledge support by NASA/JPL under NASA-Keck grant RSA \#1594531.  T. B., R. O. S., C. M.-K., and G. C. acknowledge support from the National Science Foundation, under grant number AST 1817233.  L.B. acknowledges support from NSF grant AST-1715413.  M.G. acknowledges support by the Chandra Guest Investigator program under NASA grant GO 0-21099X.  The authors would like to thank V. Ma\v cka and R.T. Gatto for their insightful discussions and support in the analysis of the work presented.  The authors would also like to thank the anonymous referee for their insightful comments and suggestions.

The data presented herein were obtained at the W. M. Keck Observatory, which is operated as a scientific partnership among the California Institute of Technology, the University of California and the National Aeronautics and Space Administration. The Observatory was made possible by the generous financial support of the W.M. Keck Foundation. The authors wish to recognize and acknowledge the very significant cultural role and reverence that the summit of Mauna Kea has always had within the indigenous Hawaiian community. We are most fortunate to have the opportunity to conduct observations from this mountain.

The scientific results reported in this article are based on observations made by the Chandra X-ray Observatory.

This publication makes use of data products from the Wide-field Infrared Survey Explorer, which is a joint project of the University of  California, Los Angeles, and the Jet Propulsion Laboratory/California Institute of Technology, funded by the National Aeronautics and Space Administration.

Funding for SDSS-III has been provided by the Alfred P. Sloan Foundation, the Participating Institutions, the National Science Foundation,  and the U.S. Department of Energy Office of Science.  The SDSS-III web site is \url{http://www.sdss3.org/.}  SDSS-III is managed by the Astrophysical Research Consortium for the Participating Institutions of the SDSS-III Collaboration including the University of Arizona, the Brazilian Participation Group, Brookhaven National Laboratory, Carnegie Mellon University, University of Florida, the French Participation Group, the German Participation Group, Harvard University, the Instituto de Astrofisica de Canarias, the Michigan State/Notre Dame/JINA Participation Group, Johns Hopkins University, Lawrence Berkeley National Laboratory, Max Planck Institute for Astrophysics, Max Planck Institute for Extraterrestrial Physics, New Mexico State University, New York University, Ohio State University, Pennsylvania State University, University of Portsmouth, Princeton University, the Spanish Participation Group, University of Tokyo, University of Utah, Vanderbilt University, University of Virginia, University of Washington, and Yale University.  

This research has made use of the NASA/IPAC Extragalactic Database (NED) which is operated by the Jet Propulsion Laboratory, California Institute of Technology, under contract with the National Aeronautics and Space Administration.

%

\vspace{5mm}


\software{astropy \citep{astropy2013},  
          TOPCAT \citep{taylor2005},
          \textit{emcee} \citep{emcee}
          }

\end{document}